# Scaling and flow profiles in magnetically confined liquid-in-liquid channels

Arvind Arun Dev[1,2], Florencia Sacarelli[3], G Bagheri[3], Aleena Joseph[1], Anna Oleshkevych[1], E Bodenschatz[4,5,6,7], Peter Dunne[1], Thomas Hermans[3]* and Bernard Doudin[1]*

[1] Université de Strasbourg, CNRS, IPCMS UMR 7504, 23 Rue du Loess, 67034 Strasbourg, France
[2] School of Applied and Engineering Physics, Cornell University, Ithaca, NY 14853, USA.
[3] Université de Strasbourg, CNRS, UMR7140, 4 Rue Blaise Pascal, 67081 Strasbourg, France
[4] Laboratory for Fluid Physics, Pattern Formation and Biocomplexity, Max Planck Institute for Dynamics and Self-Organization, Göttingen 37077, Germany.
[5] Institute for Dynamics of Complex Systems, University of Göttingen, Göttingen 37077, Germany.
[6] Laboratory of Atomic and Solid State Physics, Cornell University, Ithaca, NY 14853.
[7] Sibley School of Mechanical and Aerospace Engineering, Cornell University, Ithaca, NY 14853.

Ferrofluids kept in place by permanent magnet quadrupoles can act as liquid walls to surround a second non-magnetic inside, resulting in a liquid fluidic channel with diameter size ranging from mm down to less than 10 µm. Micro particle tracking velocimetry (µPTV) experiments and modeling show that near ideal plug flow is possible in such liquid-in-liquid channels due to the reduced friction at the walls. The measured fluids velocity profiles agree with the predictions of a hydrodynamic model of cylindrical symmetry with a minimal set of hypotheses. By introducing symmetry breaking elements in the system, we show how unique velocity and flow properties can be obtained. Our liquid-in-liquid confinement opens new possibilities for < 10 µm-sized microfluidics with low pressures and low shear, with flow characteristics not attainable in comparable solid-wall devices.

## 1. Introduction

In the realm of fluid mechanics, there is a need of new approaches to diminish friction in fluid flow [1–5] and therefore limit the transport energy loss [6,7]. A low friction environment can exist under exotic conditions like super fluidity at low temperature [8], nanofluidic channels (nanopores) made up of atomically flat crystals [9], or superfluid like behaviour of bacterial suspensions [10]. These specific physical conditions allow a velocity flow profile that approaches plug flow with a constant velocity, in stark contrast to the expected parabolic behaviour predicted by Poiseuille's flow profile constrained by no-slip boundary conditions. However, a room temperature system with low wall friction and thus low shear to transport delicate particles or cell is rare and feasible only with the use of lubricants in coaxial flows [11], or infused in channel surfaces [11–13]. The use of such lubricating layers is restricted by their limited robustness, arising from drainage of the lubricant or the occurrence of hydrodynamic instabilities [14–16]. Pioneering work in the 80s demonstrated friction reduction by trapping ferrofluid lubricants inside large pipes using magnetic forces [17,18]. In simple terms, the magnetic force holds a layer of ferrofluid in place, over which the transported liquid flows. Thus, the absence of the solid wall of the pipe leads to reduction of friction. Here we review how optimization of magnetic force fields can stabilise small fluidic tubes, aiming to scale fluidic circuity down to micro- and nanofluidic sizes. Shear forces, which are a key bottleneck for handling delicate biological objects [19], are minimised by this approach. It therefore opens the possibility to realize robust pressure-controlled microfluidic devices down to the smallest sizes.

This work is based on our initial findings of the use of a quadrupolar field to stabilize microfluidic channels and the observed large reduction of friction [20,21]. Our aim is two-fold: first specify the physical properties that govern size and friction of a transported fluid and then provide direct experimental insight by imaging a fluidic circuit and its velocity profile characteristics. We show that channels of size below 10 µm can be obtained and a nearly constant (plug-)flow velocity profile can be reached in this liquid-in-liquid design. In the last section, we propose insights into a novel fluidic behaviour that can be observed when the cylindrical symmetry of the system is perturbed.

## 2. Magnetic Force Field Design

In the presence of a non-uniform magnetic field $\vec{H}$, a material with uniform magnetization $\vec{M}$, small enough to neglect its generated demagnetizing field, experiences a Kelvin force:

$$\vec{F} = \mu_0 (\vec{M} \cdot \vec{\nabla}) \vec{H} \qquad (1)$$

where $\mu_0$ is the permeability of free space. The direction of this force depends on the magnetic susceptibility of the material, negative for diamagnetic materials and positive for paramagnetic or ferromagnetic ones, i.e. diamagnets are repulsed from regions of highest magnetic field, while the reverse is true for paramagnets and ferromagnets.

All the fluidic circuits presented here are based on the use of a pseudo-quadrupolar magnetic field source generated by commercially available N42-grade NdFeB permanent magnets of magnetization $\mu_0 M_r = 1.2$ T. These magnets are arranged to generate a negligible field at the geometric centre of the assembly while radially increasing outward. Fig. 1 shows cross sections of the two magnetic designs used for experiments, with the white arrows indicating the direction of magnetization of the magnets. Fig. 1a) illustrates the generated nearly axisymmetric magnetic field [20,21] which in cylindrical coordinates can be approximated by a magnetic field $H$ with a constant gradient:

$$\frac{\partial H(r)}{\partial r} = \frac{4 M_r}{\pi\, w} \qquad (2)$$

where the magnetic stray field $H$ is related to $B$ via $B = \mu_0 H$, $r$ is the radial distance from the centre, and $w$ the gap between magnets with the same magnetisation (Fig. 1). When a paramagnetic or superparamagnetic liquid (i.e. a ferrofluid) is inserted in the region between magnets (Fig. 1a), it is attracted to the high field regions, displacing and encapsulating any diamagnetic or weaker paramagnetic liquids which are then confined to the low field regions. The resulting liquid-in-liquid tube, or '*antitube*' [21], adapts a nearly circular magnetic cross-section.

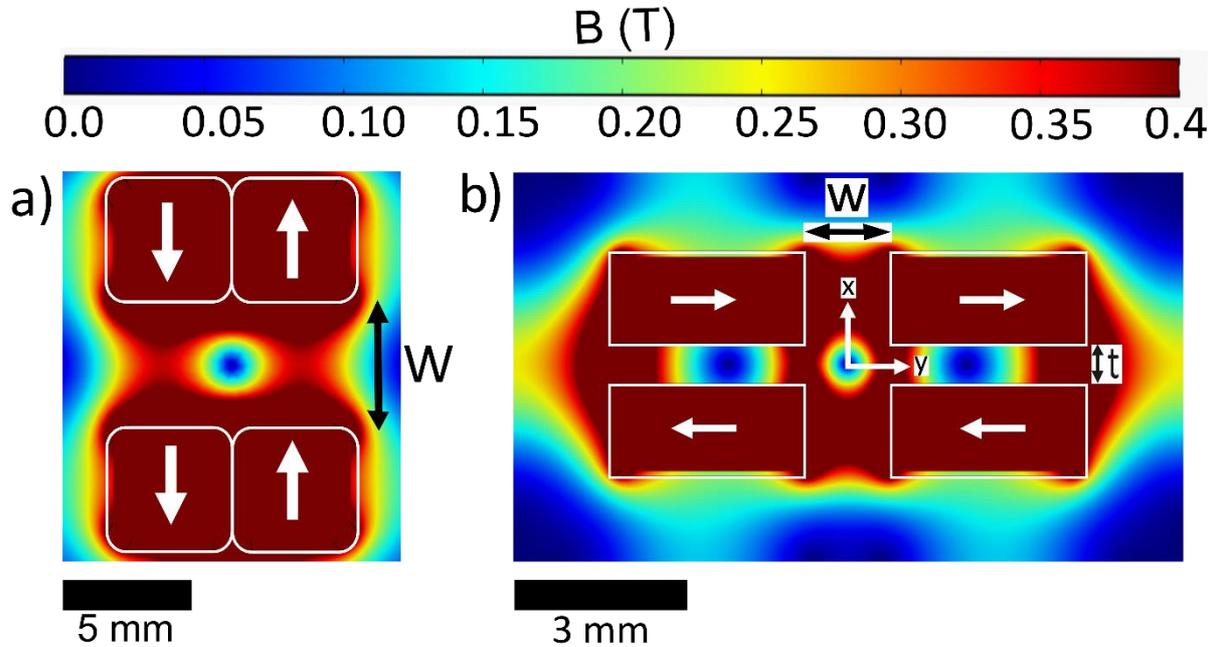

Fig. 1. Magnetic design cross section: **a)** Magnetic field generated by four magnet arrangements with white arrows showing their magnetization direction, $W = 6$ mm. The flow is in z direction (normal to the plots) **b)** Magnetic arrangement for the velocimetry experiments, with $W = 1.5$ mm, $t = 0.7$ mm.

## 3. The Static Case (No Flow)

At first let us consider a system without flow in order to test the hypothesis of a cylindrical enclosure for the transport of the liquid of interest, to determine the smallest possible diameter that can be reached. The strong optical absorption of ferrofluids complicates imaging, making the use of X-ray imaging the most appropriate tool, with synchrotron beamline facilities being necessary for micron-range resolution. Fig. 2 shows the experimental X-ray absorption contrast image using 2D radiography slices and 3D reconstruction from synchrotron tomography imaging. Fig. 2a) reveals the antitube as brighter region in the centre, of average diameter 80 ± 2 µm, surrounded by the darker ferrofluid, with Fig. 2b showing the circular cross-section of the antitube in a vertical slice. Fig. 2c) confirms the cylindrical geometry in the reconstructed data, where the antitube and surrounding ferrofluids are coloured in yellow and blue respectively.

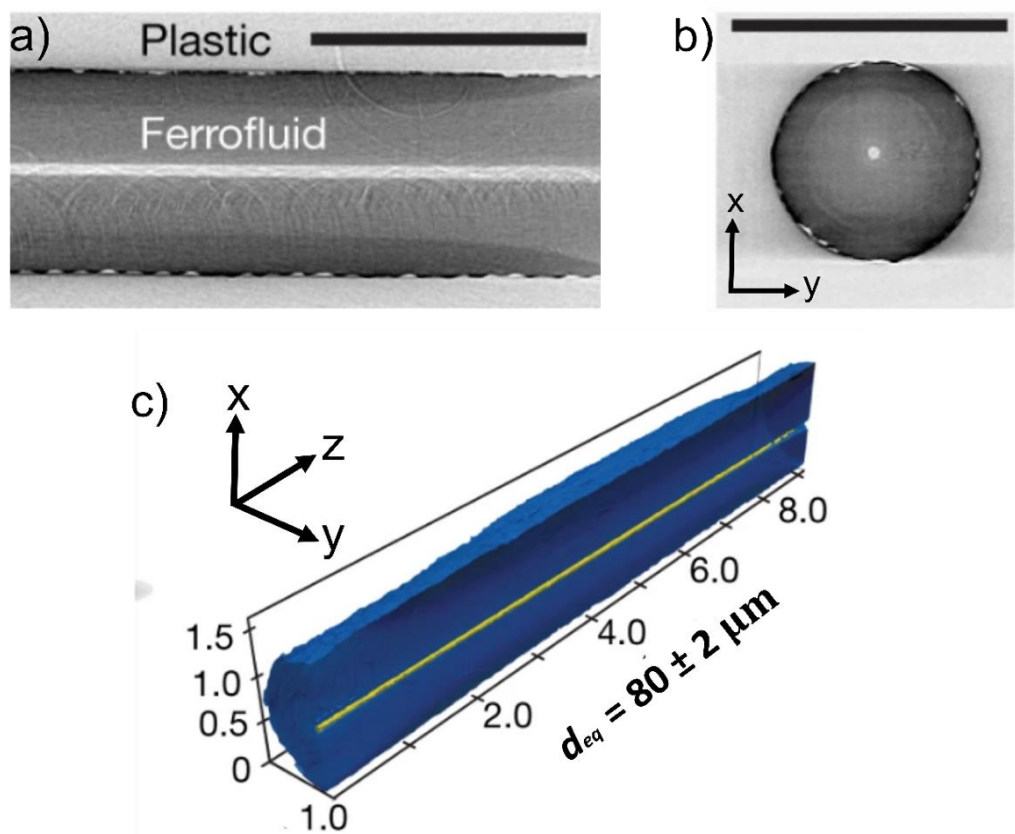

Fig. 2. Imaging of the encapsulated antitube. **a)** X-ray absorption contrast image, 2D radiography of an antitube [21], the brighter central part is water encapsulated by darker surrounding ferrofluid, **b)** cross section of Fig. 2a) where the brighter central part is the transported liquid surrounded by a darker ferrofluid. **c)** Synchrotron X-ray 3D tomographic reconstruction of a water antitube (yellow) with diameter $80 \pm 2\ \mu m$, surrounded by ferrofluid (blue). Reproduced with permission from [21].

In the static case, the balance of Laplace pressure (i.e. surface tension), magnetic pressure, and magnetic normal traction at the magnetic non-magnetic interface result in a minimum/equilibrium antitube diameter, $d_{eq}$. To begin with, the Bernoulli equation with magnetic pressure is given by:

$$P' + \rho g h + \frac{1}{2}\rho u^2 - P_m = \text{constant} \tag{3}$$

Since the magnetic force is considerably larger than the gravitational force, and we are in static case, we neglect the gravity and velocity terms giving:

$$P' - P_m = \text{constant} \tag{4}$$

Eq. 4 is the governing equation valid at all locations, where $P'$ is the local pressure and $P_m = \int_0^H \mu_0 M \, dH = \mu_0 \bar{M} H$ is the fluid magnetic pressure, $M$ is the magnetization of the ferrofluid, $\bar{M}$ is the field averaged magnetisation of the ferrofluid under external magnetic field $H$. A general boundary condition to solve Eq. 5 is:

$$P' + P_n = P_0 + P_c \tag{5}$$

where $P_0$ is atmospheric pressure. $P_n = \frac{1}{2}\mu_0 M_I^2$ and $P_c = \frac{\sigma}{R}$ are the magnetic normal traction and Laplace pressure respectively, $\sigma$ is the surface tension, and $R$ the radius of the antitube. $P_n$ and $P_c$ act at the magnetic–nonmagnetic interface and at the interface with a nonzero interfacial tension respectively. In our work, both act at the boundary between antitube and ferrofluid. Hence, at the centre of the antitube (location 1), $P_1' = P_0$ and at the magnetic-nonmagnetic interface (location 2), $P_2' = P_0 + P_c - P_n$. Since Eq. 5 is valid at all the locations we can write:

$$P_1' - P_{m_1} = P_2' - P_{m_2} \tag{6}$$

as at centre of the antitube, $H$ = 0 (see Eq. 2), hence $P_{m_1} = 0$ and Eq. 6 results in:

$$0 = \frac{\sigma}{R} - \mu_0 \bar{M} H_I - \frac{1}{2}\mu_0 M_I^2 \tag{7}$$

Eq. 7 depicts the balance between surface tension (Laplace pressure), fluid magnetic pressure and the magnetic normal traction, where $\sigma$, $H_I$ and $M_I$ are surface tension, magnetic field and ferrofluid magnetisation at the magnetic-nonmagnetic interface respectively. Rearranging Eq. 7 gives the equilibrium diameter of the antitube ($d_{eq} = 2R$) given by [21]

$$d_{eq} = \frac{4\sigma}{2\mu_0 \bar{M} H_I + \mu_0 M_I^2} \tag{8}$$

Eq. 8 shows the relative contribution of interfacial tension and magnetic properties respectively. For a fixed magnetic design, it is clear that $d_{eq}$ can be reduced by increasing the magnetic field amplitude at the interface (increased when decreasing the distance between source magnets), increasing the magnetisation of the ferrofluid, or reducing the interfacial tension. Fig. 3a) summarizes how the antitube diameter varies with magnetic separation and the choice of ferrofluid [9], with Fig. 3b) we illustrate the experimental visualization by a brightfield optical microscope of a small diameter antitube, using the magnets design of Fig 1b) and a water+surfactant antitube with MD4 ferrofluid and Tween20 as a surfactant. The annotation MD4 S denotes the ferrofluid MD4 and S the surfactant in the water antitube used to reduce the interfacial tension. Inset in Fig. 3b) show the gradual increase in ferrofluid volume to reach equilibrium diameter. Reduction in the equilibrium diameter was possible by using ferrofluid with higher saturation magnetization (EMG900) giving larger magnetic pressure. The ferrofluid EMG 900 (Ferrotec) is commercially available. Here EMG900 2S indicates EMG900 ferrofluid with a double surfactant use in water and in ferrofluid [21]. A resulting $d_{eq} = 13 \pm 0.5\ \mu m$ was obtained, shown in Fig. 3c). Reduction of $d_{eq}$, can also be obtained by decreasing the size of the cell. Bringing the magnets closer (increasing $M_I$ and $\bar{M}$) increases the magnetic pressure and further reduces the static diameter ($d_{eq}$). The experimental cell with four magnet arrangement resembles Fig. 1b) with distance between magnet pair down to 150 $\mu m$. Fig. 3d) shows the experimental image of cross sections of the antitube (bright central part) and surrounding ferrofluid EMG S. EMG S denotes the ferrofluid EMG900 with surfactant (Tween20) in water to reduce the interfacial tension.

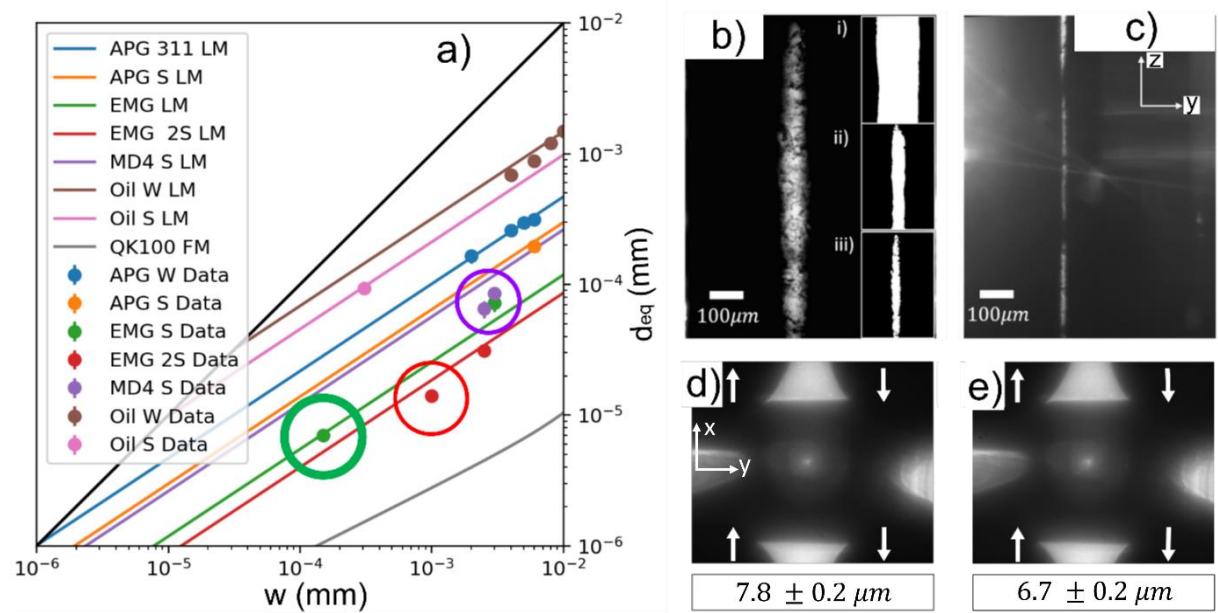

Fig. 3. Measurements of the static antitube minimum size. a) Equilibrium diameter $d_{eq}$ (Eq. 8) as a function of width W between the magnets, for several ferrofluids and antitube fluids (for abbreviation details, see [21]). b) optical microscopy image of the antitube along the (yz) plane with MD4 ferrofluid. The inset shows the gradual filling of the cavity by the ferrofluid, reaching the data points shown in purple in a) and c) image of the antitube with EMG900 ferrofluid with added surfactant and water antitube with added surfactant, in the so-called double surfactant (EMG 2S) configuration, with data point in red in a). X-ray tomography circular cross sections shown in d) and e) are (xy) cut images of a water antitube in EMG900 ferrofluid with added surfactant, (single surfactant EMG S), with related green encircled data point indicating the smallest measured of 6.7 $\mu m$ in a). Fig 3a is modified from the figure published in [21], with permission. Fig 3d and fig 3e were obtained using the X02DA TOMCAT X-ray beamline of Swiss Light Source (SLS) at the Paul Scherrer Institute, Villigen, Switzerland.

The minimum diameter of the antitube is 6.7 $\pm$ 0.2 μm, measured using synchrotron X-Ray tomography with 0.1 μm resolution. The antitube diameter data point is presented in Fig. 3a) along with the data obtained by Dunne et. al. [21] Our data is encircled (green) in Fig. 3a) along with the data corresponding to Fig. 3a) and Fig. 3b). Experiments agree well with the model curve [21]. LM and FM denotes the linear model and full model respectively. The linear model, LM ($M = \chi H$) is valid at small magnetic field whereas the full model considers the nonlinearity in the ferrofluid magnetization curve [21]. Note that antitubes larger than $d_{eq}$ are always possible to make, by injecting less ferrofluid volume. As such, $d_{eq}$ is a lower limit of practical use for the antitubes.

In summary, the stability of such cylindrical antitube is governed by the relative strengths of the destabilizing Laplace pressure and the stabilizing magnetic pressure. Since, the Laplace pressure is given by the ratio of interfacial tension and diameter of the flow channel, increasing when the size of the fluidic circuit decreases. It therefore sets the limit for smallest achievable diameters when it exceeds the magnetic pressure. The latter depends on the magnitude of the applied local field and the magnetic susceptibility of the ferrofluid material. A suitable design of the magnetic field and the appropriated choice of intrinsic properties of the ferrofluid make possible the stabilization of minimal antitube size ($d_{eq}$). In a nutshell, one needs to maximize the magnetic susceptibility of the ferrofluid and minimize its interfacial surface tension with the enclose liquid forming the antitube fluidic circuit. Our experiments show that diameters below 10 μm can be realized, with possible smaller values requiring further miniaturization of the source magnets (for increasing the local magnetic field) and optimization of the ferrofluid intrinsic properties. Fig. 3 indicates that the latter is possible, but imaging resolution issues make the observations of diameters below 1 μm elusive.

## 4. The Dynamic Case (Under Flow)

In the dynamic case, we detail the hydrodynamics resulting from the motion of the transported liquid. The conceptual difference between the standard flow (Poiseuille flow) and the proposed flow (antitube flow) is illustrated in Fig.4.

A standard Poiseuille flow in Fig.4 (top) is defined as a viscous flow with zero velocity at the confining wall (no-slip condition). This gives the well-known parabolic velocity profile with maximum velocity at the centre of the tube. In the proposed antitube flow with ferrofluid lubrication, the confining wall is a liquid (ferrofluid) shown in Fig. 4 (bottom). Since at the liquid-liquid interface, the velocity has to be unique (no slip condition), the confining liquid (ferrofluid) also moves along with the transported liquid. This makes lubricated flow, a system with moving wall. For small enough flow rates, the magnetic force field holds the ferrofluid in place, avoiding its shearing out. The resulting volume conservation of ferrofluid implies a recirculation of the ferrofluid, depicted by the reverse flow in the darker region of the ferrofluid in Fig. 4. This proposed flow design results therefore in:
 1. a finite velocity at the confining wall
 2. a reverse flow path of the ferrofluid.

These two distinct features lead to spectacular high drag reduction values, measuring the reduction of friction forces, up to measured values of 99.8% for viscous liquids [20], or nearly frictionless flow. Fig. 4 (middle) compares the flow velocity in the cross section for Poiseuille flow (left) and antitube flow (right). For the antitube, the velocity of flow is of nearly uniform amplitude across the flow channel, depicting near zero shear. The velocity profile (arrows) presented in Fig.4 bottom is a result of solving the Navier-Stokes (N-S) equation with appropriate boundary condition, under the hypothesis of non-deformation of the cylindrical liquid-in-liquid flow, as observed experimentally for low enough flow rates.

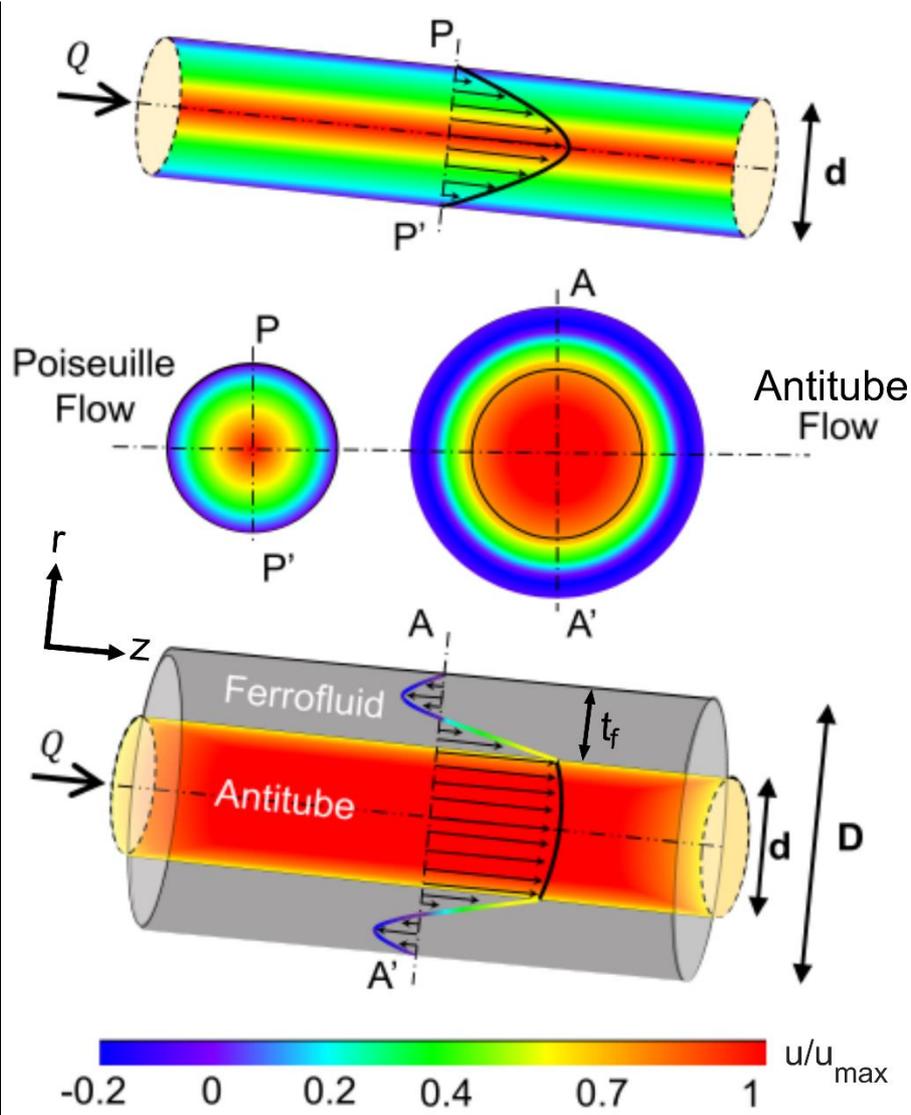

Fig. 4. Differences between Poiseuille flow and magnetically confined flow channels. Top: Poiseuille flow through a tube of diameter d, with flow rate Q, and the parabolic velocity profile showing zero velocity at the flow wall. Bottom: Magnetically confined flow channel (antitube) with transported liquid in antitube and ferrofluid encapsulation. The flow wall (between antitube and ferrofluid) has finite velocity. Ferrofluid exhibits a counter flow circulation. D is the flow cavity. Middle: Comparison of cross section of flow for the two flow systems. Almost constant velocity (ideal flow) for magnetically confined flow compared to a uniform velocity gradient over the cross section for Poiseuille flow.

In Fig. 4 (bottom), the interface between antitube (bright central part) and ferrofluid (darker encapsulation) is the liquid-liquid interface. The outer boundary of the ferrofluid is the solid boundary of the plastic cavity. A flow rate $Q$ through a channel of width $d$ is imposed for the transported liquid with the ferrofluid lubrication thickness being $t_f$. $D$ is the width of the microchannel cavity. Since the Reynolds number is small (<1), we can neglect the inertial terms in the N-S equation and the flow is viscous dominated, modelled and explained by the Stokes equation [20]. The wall velocity is given by [20]:

$$\frac{u_{wall}}{u_{max}} = \frac{\beta_0 - 1}{\beta_0 + 1} \quad (8)$$

$$\beta_0 = 1 + 4\eta_r \ln(1 + 0.8 t_f^\star) \quad (9)$$

where, $\beta_0$ is the simplified drag reduction factor [20] and $\eta_r$ is the viscosity ratio of the two fluids. The normalized thickness of ferrofluid is described as $t_f^\star = \frac{t_f}{d}$. The wall velocity depends on the relative coverage of the ferrofluid ($t_f^\star$) and the viscosity ratio ($\eta_r$). As expected, increaing the relative amount of ferrofluid improves the lubrication. Inversely, if the viscosity of the ferrofluid increases, the systems is more and more resembling to a solid wall boundary condition, therefore detrimental to lubrication.

Velocimetry experiments require no obstruction to the incident light (no ferrofluid in the path of light), hence a planar flow equivalent, slightly deviating from axisymmetric encapsulation of the transported liquid, is required. In this case, the ferrofluid covers the side of the flow channel, keeping the light unobstructed (See Fig. 5c).

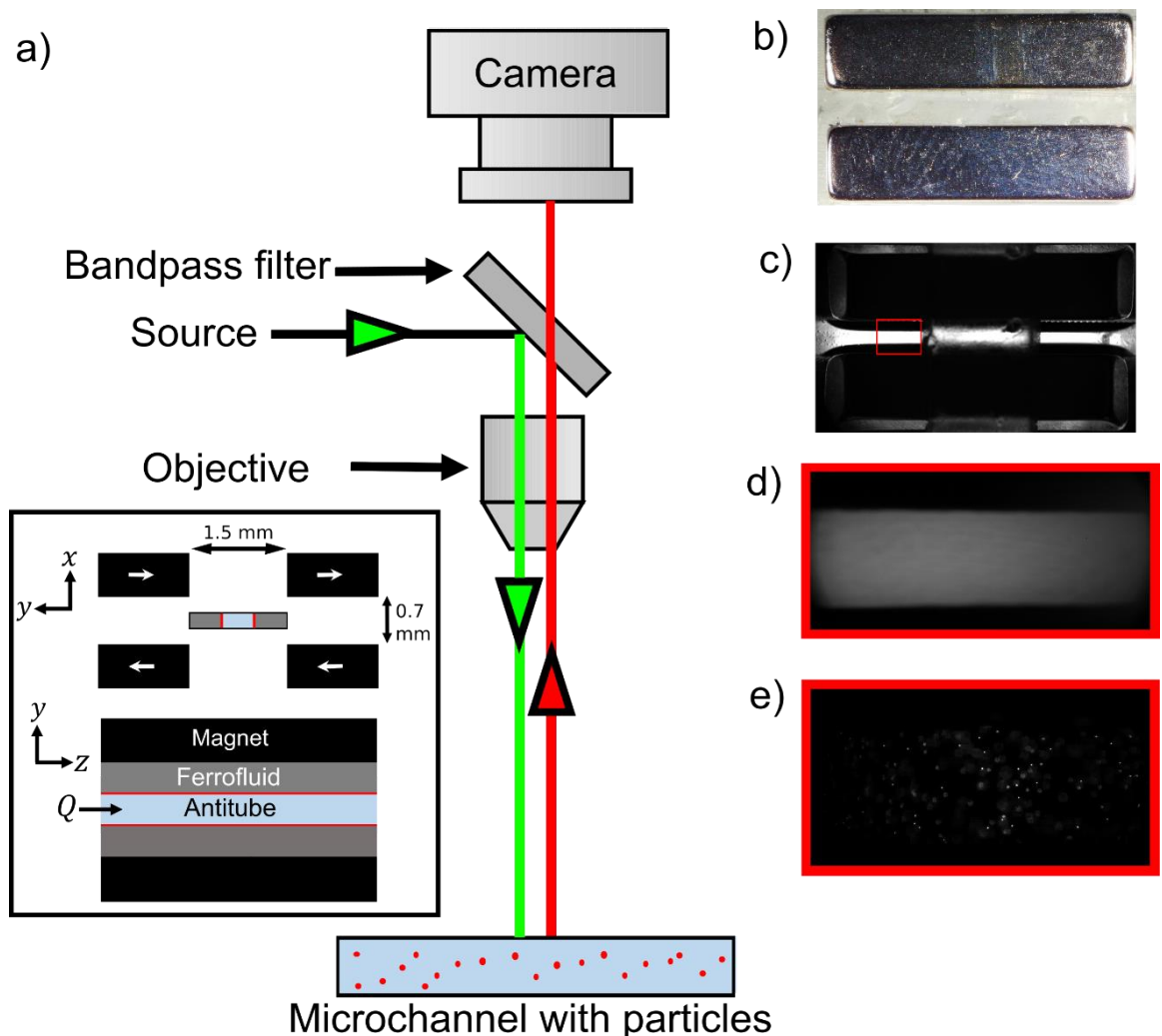

Fig. 5. Experimental set up and direct measurement of velocity profile. a) Optical system for µPTV measurements, b) microfluidic channel without ferrofluid, c) microfluidic channel with ferrofluid, d) brightfield image of region of interest under microscope, e) fluorescence particles signal. Inset in Fig. 5a) shows the schematic of microchannel and magnetic arrangement.

We restrict ourselves to measure the flow profile away and equidistant from the two non-lubricated surface to minimize the solid walls influence. Inset in Fig. 5a) show the schematic of the magnetic and flow arrangement, with a cross section view of the four magnets in the quadrupolar arrangement and a top view showing the imposed flow rate $Q$ along the length of the microchannel. The magnetic field generated by the arrangement is shown in Fig.1b). The four white arrows in the cross section show the direction of magnetization of the magnets and the red line along the flow depicts the liquid-liquid interface. As required, the ferrofluid (colour grey) only covers the two walls and hence facilitates the transmission of light, essential for flow visualization.

We measure the velocity profile in the antitube directly using μPTV. We use glycerol ($\eta_a$ = 1.1 Pa.s) as antitube transported liquid and ferrofluids APGE32 ($\eta_a$ = 1.7 Pa.s) from Ferrotec [22] and a biocompatible ferrofluid ($\eta_a$ = 0.144 Pa.s) from Qfluidics [23] for testing the flow behavior with different viscosity ratios, $\eta_r = \frac{\eta_a}{\eta_f}$. The antitube and the microchannel cavity width are $d$ = 0.5 mm and $D$ = 1.5 mm respectively. The ferrofluid coverage is maintained at, $t_f = \frac{D-d}{2}$. We begin by mixing glycerol with fluorescence 4 $\mu$m size particles (FluoSpheres™). The excitation and emission maxima for the particles are 580 nm and 605 nm respectively. We use a source of light with a bandpass filter (572/25 for excitation and 629/62 for emission). Fig. 5a) shows the schematic of flow setup used for the velocimetry measurements. Fig. 5b) and Fig.5c) show the microfluidic channel before and after inserting the ferrofluid. Fig. 5d) shows a magnified brightfield image of the region of interest for PTV measurements and Fig. 5e) shows particles under fluorescence in the micro channel.

The depth of field at this magnification is 3 $\mu$m, below the chromophores diameter (4 $\mu$m). This makes possible recording images with minimum contribution from the off-focus particles (below or above the imaging plane). The time elapsed image frames are recorded with Zeiss Axio zoom V16 microscope and a Phantom v2511 camera. The exposure time is 400 μs. The images are analysed by tracking differences in positions of particles in consecutive frames using FIJI TrackMate [22]. Data is then analysed in Python to bin the data for different heights and calculate the standard error for each bin height. Fig.6a) and Fig.6b) show the analytical Poiseuille flow and antitube experimental velocity profile using μPTV measurements. Markers are experimental and solid line stands for predicted analytical velocity profile [20]. Black lines denote the Poiseuille flow. The errors correspond to the standard deviation of velocities measured over 10000 image sequences. The antitube lies between y/d = [− 0.5 0.5]. The Poiseuille flow shows the zero velocity at the wall coordinates, y/d = 0.5 and y/d = −0.5. Experimental markers in Fig.6a) show that for $\eta_r$ = 0.65 with APGE32 as ferrofluid, the wall velocity at the liquid-liquid interface is large enough, up to 60% of the maximum velocity. This wall velocity reaches almost 85% of the maximum velocity for $\eta_r$ = 7.64 as seen in Fig.6b) for a bio-compatible ferrofluid. The ferrofluid coverage is slightly asymmetric for $\eta_r$ = 0.65, as the ferrofluid thickness are a bit different between left and right. For $\eta_r$ = 7.64, the ferrofluid coverage is symmetric. The increase of wall velocity with increase in $\eta_r$ is also forecasted by the analytical predictions.

Fig.6a) and Fig.6b) show direct evidence of large wall velocity with a magnetic fluid lubrication as previously predicted by some of us [20,21]. The increase of wall velocity with increase in $\eta_r$ is also foreseen by the analytical predictions and relates to the expected enhanced lubrication. The wall velocity values predicted for APGE32 (fig.6a)) and the bio compatible ferrofluids (fig.6b)) given by Eq.8 are 32% and 90% of the centre velocity maximum respectively.

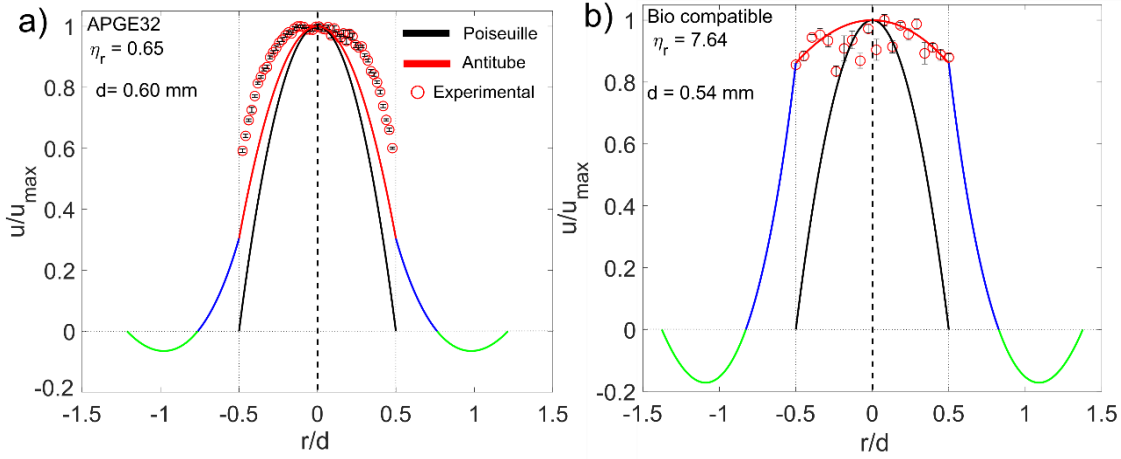

Fig. 6. Comparison of experimental velocity profiles (markers) with Poiseuille flow (Black line). a) For APGE32 ferrofluid with viscosity ratio $\eta_r$ = 0.65 and b) for biocompatible ferrofluid with $\eta_r$ = 7.64. Lines are the analytical model solution of Stokes equation, red line for the antitube, green and blue line for the ferrofluid.

The experimental and analytical predictions agree with some deviations. The experimental measurements might differ from that theorized in literature [20] due to presence of two solid walls, which is not taken into account in the analytical modelling [21]. Nonetheless, the direct measurements do confirm the presence of large wall velocity at the liquid-liquid interface which were only indirectly claimed earlier [20,21]. We have limited ourselves here to the regime of low flow rate, to avoid capillary instabilities that can develop at higher flow rates. This may result in deformations in the magnetic-nonmagnetic interface, large enough to result in shearing of the ferrofluid or shear failure of the lubricated flows. More sophisticated analytical approach and numerical simulations are necessary to extend this simple model to include liquid wall deformation or to understand flow behaviour at the extremities of the circuit, beyond the scope of the present work. Instead, we focus on testing how solid cavity design changes can impact the antitube flow profile, with an example discussed in the next section.

## 5. Beyond a simple cylindrical symmetry

We argue that the symmetry of the flow can be broken by implementing deformations all along the chamber walls (i.e., only in contact with the magnetic fluid). One possible implementation is to define a threaded screw profile along the cylindrical circuit length (Fig. 7a)), where a small rectangular section was left free, to allow for proper imaging using confocal microscopy. The magnetic design is given in the inset of Fig. 7a). Fig. 7b) illustrates the occurrence of a vortex flow profile resulting from the screw-shaped solid walls boundary conditions. The prediction of fluid motion with complex fluid interfaces poses a challenge. Here, we qualitatively explain our experimentally observed flow in screw wall device using a forced vortex model. A forced vortex in two dimensions (2D) implies the occurrence of an azimuthal velocity $V_\theta^\star$ of simplest expression [25]:

$$V_\theta^\star = \omega^\star r^\star \qquad (10)$$

stating it is proportional to the radial position $r^\star = \sqrt{x^2 + y^2}$ through the angular velocity $\omega^\star$, with all variables non-dimensionalized by their values at the magnetic-nonmagnetic interface (flow wall).

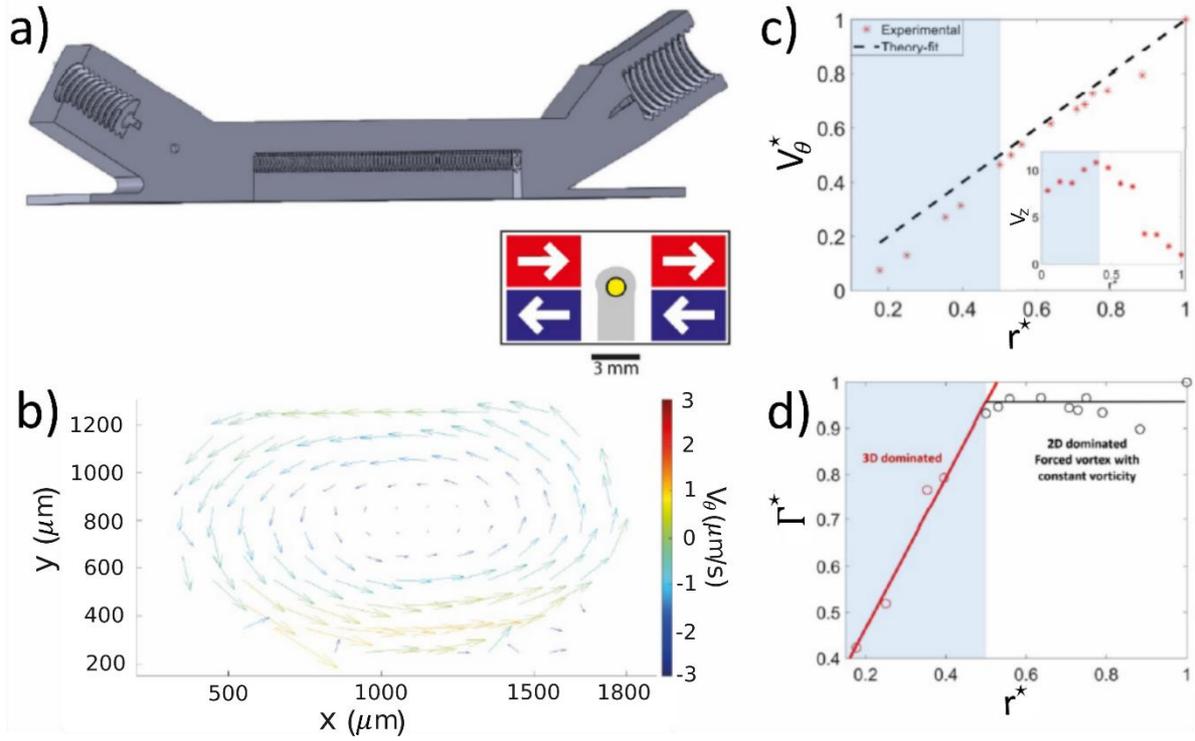

Fig. 7. Vortex flow in antitube. a) Screw walled device for vortex flow design, inset shows configuration of 4 magnets in the design with yellow central part depicting the antitube, b) vortex flow observed by velocimetry, c) forced vortex model and experimental data, inset show the axial velocity ($V_z$) in the direction of flow with respect to the radial coordinate and d) experimental vorticity $\Gamma^\star(= 2\omega^\star)$ as a function of reduced radial length.

Fig. 7c) shows the azimuthal velocity with respect to radial coordinate. It can be seen from Fig. 7c) that the experimental data obtained for azimuthal velocity matches well with the theoretically predicted fit for $r^\star > 0.5$, with flow similar to a forced vortex. The inset in Fig. 7c) shows that the axial velocity ($V_z$) (in the direction of flow) decreases sharply after $r^\star > 0.45$. This reduced axial velocity makes a 2D model, implicit in the Eq. 10, relevant. Indeed, a key signature of a forced 2D vortex flow is the constant vorticity given by, $\Gamma^\star = 2\omega^\star$. Fig. 7d) shows the vorticity plot, illustrating the transition from a 3D to a 2D behaviour when $r^\star$ reaches 0.5. The blue highlighted region in Fig. 7c) and in the inset show the change in behaviour of azimuthal velocity and axial velocity respectively with $r^\star$, confirming the change of behaviour at $r^\star \approx 0.5$ highlighted in Fig. 7d). We therefore find that the flow has two distinct behaviours; the flow is 2D dominated when $r^\star > 0.5$ (lower axial velocity) and possibly 3D augmented for $r^\star < 0.5$.

## 6. Conclusions

Ferrofluid encapsulation of liquid circuits opens new possibilities for microfluidics and fluidics applications [21]. Stabilizing an encapsulated cylindrical 'antitube' of dimensions below 10 µm is achieved. This size value is governed by the liquid-liquid interface surface tension, as well as the scale of the surrounding assembled magnetic force field sources and the magnetic properties of the ferrofluid. Values below 1 µm are possible, however very challenging to image. Magnetic forces keeping the lubricating magnetic liquid in place mitigate the fundamental issue of stability of a liquid-in-liquid flow [15]. This makes possible to extend achievable drag reduction, resulting in remarkably large values, possibly exceeding 99% [20]. A direct measurement of velocity profile using the $\mu$PIV confirms the occurrence of liquid-liquid wall velocity as large as 85% of the maximum velocity. This value increases with the increase in viscosity ratio between transported liquid to the lubricant. The experimental observations are supported and predicted by an analytical modelling using the Stokes equation. Such low drag flow channels enable microfluidics applications that require reduced operating pressures, specifically needed in microchannel flow of concentrated solution or variable viscosity solutions. Drug delivery and shear control on delicate cells also falls in the gamut of possible applications. The magnetically confined flow channels open therefore new possibilities in the field of microfluidics for shearless transport, for a wide range of biological and technological applications. Beyond a simple flow profile (Poiseuille like parabolic profile), complex flow profiles can be obtained by tuning the design of the cavity used for the flow. We have shown that a screw pattern on the inner boundary of the cavity results in the swirling motion of flow about the central axis. The flow resembles a forced vortex flow with constant vorticity and deviates as we approach the centre of the flow, away from the screw design. These examples provide insight into the possibility to design and investigate novel flow patterns, otherwise not possible when using the solid walls that constrain the flow of the transported liquid.

**Acknowledgements**
This project has received funding from the European Union's Horizon 2020 research and innovation programme under the Marie Skłodowska-Curie grant agreement MAMI No 766007 and QUSTEC No. 847471. We also acknowledge the support of the University of Strasbourg Institute for Advanced Studies (USIAS) Fellowship, the Fondation Jean-Marie Lehn and the Interdisciplinary Thematic Institute QMat, as part of the ITI 2021 2028 program of the University of Strasbourg, CNRS and Inserm, was supported by IdEx Unistra (ANR 10 IDEX 0002), and by SFRI STRAT'US project (ANR 20 SFRI 0012) and EUR QMAT ANR-17-EURE-0024 under the framework of the French Investments for the Future Program. We acknowledge the Paul Scherrer Institut, Villigen, Switzerland for provision of synchrotron radiation beamtime at the TOMCAT beamline X02DA of the SLS and would like to thank Dr. Anne Bonnin for invaluable assistance."